\documentstyle[sprocl]{article}

\def\Journal#1#2#3#4{{#1} {\bf #2}, #3 (#4)}

\def\NPB{{\em Nucl. Phys.} B}
\def\PLB{{\em Phys. Lett.}  B}
\def\PRL{\em Phys. Rev. Lett.}

\def\be{\begin{equation}}
\def\ee{\end{equation}}
\def\bea{\begin{eqnarray}}
\def\eea{\end{eqnarray}}

\newcommand{\beq}{\begin{equation}}
\newcommand{\eeq}{\end{equation}}
\newcommand{\beqa}{\begin{eqnarray}}
\newcommand{\eeqa}{\end{eqnarray}}

\begin{document}
 \begin{flushright}
  \begin{tabular}{l}
  SLAC-PUB-7272\\
  hep-ph/9609215\\
  August 1996
  \end{tabular}
  \end{flushright}
  \vskip0.5cm

\title{POWER CORRECTIONS, RENORMALONS AND RESUMMATION~\footnote{
Research supported 
by the Department of Energy under contract DE-AC03-76SF00515.}
}

\author{ M. BENEKE }

\address{Stanford Linear Accelerator Center,\\ 
Stanford University, Stanford, CA 94309, U.S.A.}

\vspace{1.5cm}

\maketitle\abstracts{
I briefly review three topics of recent interest concerning 
power corrections, renormalons and Sudakov resummation: (a) 
$1/Q$ corrections to event shape observables in $e^+ e^-$ 
annihilation, (b) power corrections in  Drell-Yan production 
and (c) factorial divergences that arise in resummation of large 
infrared (Sudakov) logarithms in moment or `real' space.}

\vspace*{2cm}

\begin{center}
{\em To appear in the Proceedings of the \\
     28th International Conference\\
     on High Energy Physics,\\
     Warsaw, Poland, 25-31 July 1996}
\end{center}

\newpage

\section{Utilizing renormalons}

Perturbative QCD relies on factorization. By this one implies 
that an observable that depends on at least one hard scale $Q$ 
can be expanded in powers and logarithms of $\Lambda/Q$, where 
$\Lambda$ is the intrinsic QCD scale. At least up to some order 
in $1/Q$, one must also be able to factor a `short-distance' 
part from long-distance contributions, which are independent of 
the details of the hard process. At leading order in $1/Q$, the 
long-distance contribution can be absent, like in $e^+ e^-$ 
annihilation, or a product of parton distributions like for 
inclusive quantities in hadron-hadron collisions. Beyond leading 
order, little is known about power corrections, with exceptions 
like deep inelastic scattering. For event shapes 
or effects referred to as `hadronization', it is not known 
how to express 
power corrections in terms of operators and asymptotic states. 

The renormalon approach to power corrections uses the fact that the 
leading term in the power expansion already indicates the existence 
of power corrections, because the perturbative expansion of its 
short-distance coefficient diverges. This `renormalon' divergence 
occurs, because certain higher-order diagrams (the simplest being 
`bubble graphs') contain many powers of logarithms of a loop 
momentum, which make these diagrams sensitive to large 
distances.\footnote{The logarithms also enhance the sensitivity 
to distances much shorter than $1/Q$. The corresponding ultraviolet 
renormalons will not be discussed here.} Summing a divergent 
series requires a prescription. The prescription-dependence is best 
captured by the ambiguity of the Borel integral and takes the 
form $(\Lambda/Q)^a \ln^b Q/\Lambda$, where infrared (IR) renormalons 
yield positive integers for $a$. One can interpret the 
ambiguity as an ambiguity in separating long- and short-distances, 
much as the factorization scale dependence in separating coefficient 
functions and parton distributions. Since the physical observable 
is unambiguous, the ambiguity in defining perturbation theory 
must be matched by power corrections and this determines their 
$Q$-dependence, but not their magnitude, just as the evolution of 
parton distributions is perturbatively calculable, but not their 
initial values. 

The advantage of the method is that it is entirely perturbative, 
although to all orders. To some extent, the language inherited 
from studies of large-order perturbation theory is an 
unnecessary complication, since the set of diagrams 
that leads to a divergent series 
really only probes the IR sensitive regions of low-order 
skeleton-like graphs. It would be desirable to classify these 
regions systematically by extending standard methods 
of perturbative factorization \cite{fact} that identify logarithmic 
IR divergences to subleading, power-like IR sensitivity. Meanwhile, 
most calculations are done in the formal $N_f\to\infty$-limit, which 
selects diagrams with one chain of fermion loops at leading order. 
Provided the phase space of a cut fermion loop is integrated 
unweighted, the structure of power corrections inferred from renormalons in 
this approximation is equivalent to calculating the low-order diagrams 
with finite gluon mass $\lambda$ and interpreting non-analytic terms 
in the small-$\lambda^2$ expansion as power corrections.\cite{BBZ} 

The method has its limitations, precisely because it is purely 
perturbative. To go beyond classifying the expected power corrections 
for each observable separately, one needs additional assumptions, which 
do not follow from perturbative considerations alone,
such as universality of power corrections, to relate different 
observables. And, of course, the IR sensitivity of Feynman diagrams 
might not exhaust all possibilities for power corrections. 

In the past two years, these ideas have been applied to observables 
that do not admit an operator product expansion, such as event 
shapes in $e^+ e^-$ annihilation, jet observables and the Drell-Yan 
cross section. In these cases, IR renormalons provide genuinely new 
information about power corrections. 
This talk gives a somewhat qualitative overview and 
emphasizes the outstanding issues.

\section{Power corrections to event shape variables}

Event shapes are constructed from IR safe weights on hadronic final 
states in $e^+ e^-$ collisions. Thrust, for example, is defined as 
$T=\mbox{max}_{\vec{n}} (\sum_i |\vec{p}_i\cdot\vec{n}|)/\sum_i 
|\vec{p}_i|$. The theoretical prediction is computed in terms of 
parton momenta, while hadron momenta are measured. Matching partons 
and hadrons is dealt with as a hadronization correction, which is 
obtained from Monte Carlo programs and accounted for in determinations 
of $\alpha_s$ from event shapes. The fragmentation models built into 
Monte Carlo programs lead to hadronization corrections that vary like 
$1/Q$ with the cms energy $Q$. 

Theoretically one considers hadronization as a soft parton phenomenon 
that takes over from the parton shower at a certain typical hadronic 
scale $\mu_h$. This separation scale is not uniquely fixed and the 
boundary between perturbation theory and hadronization is vague. 
Thus, probing the boundary of perturbation theory with renormalons 
may tell us more about hadronization. Event shape variables have been 
computed to this end both with a finite mass gluon in the lowest order 
gluon emission diagram \cite{WEB94,DOK95,AKH} and in the approximation 
of a single chain of fermion loops \cite{NAS95}, in which case the 
region of small invariant mass of the $q\bar{q}$ pair in a cut fermion 
loop is the important one. The two methods are not equivalent in this 
case, because the invariant mass distribution depends on how each 
particular event shape weights the $q\bar{q}$ phase space.

The calculation of the average $\langle 1-T\rangle$ with finite gluon 
mass leads to a $1/Q$-correction
\begin{equation}
\langle 1-T\rangle _{|1/Q} = K\cdot\frac{\lambda}{Q}.
\end{equation}
The $1/Q$-correction comes only from the two-jet region $T\to 1$, when 
the gluon momentum becomes very small. Multiple gluon emission diagrams 
modify the constant $K$, but only if all gluons are soft, so that again 
$T\to 1$. Consequently, if the two-jet region is excluded from the average 
over $T$, we expect a smaller hadronization correction,
\begin{equation}
\frac{\langle 1-T\rangle_{|1/Q,0.5<T<0.8}}{
\langle 1-T\rangle_{|1/Q,
\mbox{\footnotesize all}\,T}} \propto \alpha_s(Q)\,.
\end{equation}
The same conclusion applies to the heavy jet mass average $\langle M_h^2
\rangle$. The DELPHI collaboration has reanalyzed \cite{delphi} the 
energy dependence of event shapes by adding $1/Q$ and $1/Q^2$ terms to 
the next-to-leading order perturbative expression (evaluated at 
scale $\mu=Q$) and by fitting the coefficients of the power corrections 
to the data. No hadronization correction from Monte Carlo programs 
is applied. Some of the results are reproduced in Tab.~\ref{table1} 
and agree qualitatively with the above predictions. The energy-energy 
correlation (EEC) is predicted~\cite{NAS95} to have $1/Q$-corrections 
at all angles, because the soft gluon region contributes at all angles. 
It is also easy to see that the three-jet rates $R_3$ computed from the 
JADE clustering algorithm have $1/Q$-corrections, while the Durham 
algorithm should have only $1/Q^2$-corrections, because it weights the 
region of soft gluons quadratically with their energy rather than 
linearly. The DELPHI analysis does not have enough data points to 
test this prediction for the Durham algorithm.

\begin{table}[t]
\addtolength{\arraycolsep}{-0.01cm}
\renewcommand{\arraystretch}{1.3}
\caption{\label{table1}
Fits to the $Q$-dependence of event shape variables. $\alpha_s(M_Z)$, 
the coefficient of a $1/Q$-term, $C_1$, and $1/Q^2$-term, $C_2$ 
(not quoted) are 
fitted to obtain the second entry for each observable. For the first 
entry $C_2$ is fixed to zero. }
$$
{\small
\begin{array}{|c||c|c|c|}
\hline
\mbox{Observable} & C_1/\mbox{GeV} & \alpha_s(M_Z) \\ 
\hline\hline
\langle 1-T\rangle & 0.82\pm 0.07 & 0.123\pm 0.002 \\[-0.1cm]
   & 0.83\pm 0.20 & 0.122\pm 0.004 \\ \hline
\int_{0.2}^{0.5}d T\,(1-T) & 0.37\pm 0.05  & 0.121\pm 0.008 \\[-0.1cm]
   & 0.20\pm 0.05 & 0.134\pm 0.003 \\ \hline
\langle M_h^2/E_{vis}^2 \rangle & 0.54\pm 0.08 & 0.121\pm 0.002 \\[-0.1cm]
   & 0.75\pm 0.26 & 0.116\pm 0.006 \\ \hline
\int_{0.1}^{0.4} \!dM_h\,(M_h^2/E_{vis}^2) & -  & - \\[-0.1cm]
   & -0.01\pm 0.03 & 0.123\pm 0.000 \\ \hline
\langle M_d^2/E_{vis}^2 \rangle & 0.19\pm 0.04 & 0.094\pm 0.003 \\[-0.1cm]
   & 0.10\pm 0.05 & 0.097\pm 0.003 \\ \hline
\int_{-0.5}^{0.5} d\cos\Theta\,EEC & 1.68\pm 0.05 & 0.115\pm 0.002 \\[-0.1cm]
   & 0.27\pm 0.23 & 0.137\pm 0.004 \\ \hline 
R_3^J(y_{cut}=0.08) & 0.44\pm 0.15 & 0.107\pm 0.001  \\[-0.1cm] 
   & -3.59\pm 0.55  & 0.123\pm 0.002 \\ \hline
R_3^D(y_{cut}=0.04) & -0.67\pm 0.49 & 0.126\pm 0.004  \\[-0.1cm]
   & -2.53\pm 3.15  & 0.137\pm 0.019  \\ \hline
\end{array}
}
$$
\end{table}

The examination of expected power corrections provides some guidance to 
selecting `good' event shapes, the good ones being those less sensitive 
to hadronization. To go further, one has to make the stronger assumption 
that hadronization corrections in the two-jet limit are 
universal.\cite{DOK95,AKH} This implies that although the constants
$K$ above are not calculable for any observable, their ratio for 
different observables is calculable, because multiple soft parton emission 
modifies $K$ in a universal way. Thus, fitting a $1/Q$-correction to 
one observable would determine the hadronization parameter 
once and for ever. 
The assumption of universality could 
be justified diagrammatically if an event shape variable did not resolve 
the soft parton kinematics, which in fact it does. \cite{NAS95} For 
example, in the 
two-jet region $1-T\approx (M_h^2/Q^2)+(M_l^2/Q^2)$, where $M_l$ is the 
light jet mass. If we now consider the diagram where a single emitted 
gluon splits into a $q\bar{q}$ pair, we find $1-T=M_h^2/Q^2$ if both 
quarks are emitted into the same hemisphere, and $1-T=2 M_h^2/Q^2$ 
if they are emitted into opposite hemispheres. Thus, there is no 
unique relation between $1-T$ and $M_h^2$, even if both quarks are 
soft. Universality could still hold in an approximate sense, if, 
as advocated in Ref.~\cite{DOK95}, the 
strong coupling approaches a finite and not too large value in the 
infrared. In this case, the diagram just discussed is higher order 
in the IR coupling. In this scenario, the $1/Q$-correction to 
$M_l^2$ should be smaller than for $M_h^2$, because it arises only 
at second order. The small power correction for the average 
$M_d^2=M_h^2-M_l^2$ does not support this picture, although the small 
fit value for $\alpha_s(M_Z)$ indicates that the corresponding 
$C_1$ in Tab.~\ref{table1} might not be too reliable.

Eventually, universality should be subjected to 
experimental tests. In this respect, it would be interesting to obtain 
the coefficients $C_1$ in Tab.~\ref{table1} with $\alpha_s$ fixed 
to a unique value. As a matter of principle, the power corrections 
obtained by renormalon methods are synonymous with large perturbative 
corrections in higher orders. If large coefficients are a practical 
concern, the divergent piece of the series should be separated and 
discarded, so that it is absorbed into the power correction, leaving 
an unambiguous perturbative series. A procedure of this sort has been 
proposed in \cite{DOK95} and applied with some success to $\langle 1-T
\rangle$, the average $C$-parameter and $\sigma_L$. Another question of 
importance for testing universality is to what extent the power 
corrections in Tab.~\ref{table1} effectively parameterize higher 
order corrections in perturbation theory, which would in principle 
be calculable, leaving a rather small `true' hadronization correction. 
In Ref.~\cite{proc} it was argued that higher order corrections, 
summed up to the point where the series diverges, can well mimic the 
shape of a $1/Q$-correction. An equivalent effect is obtained, if 
one expresses the second order perturbative prediction in terms 
of $\alpha_s(Q^*)$ with $Q^*\sim 0.1 Q$. Such a low scale is not 
unnatural for event shape observables, since they are dominated by 
the soft-collinear region, where the scale is set by the transverse 
momentum of the emitted gluon rather than $Q$. From this point of 
view, the question of whether universality holds is less important 
than determining the higher-order perturbative corrections or correct 
scale for each event shape.

In principle, the universality assumption could also be invoked 
to relate two non-IR safe event shapes to each other. 
This would circumvent the difficulty of having to extract subdominant 
power corrections to test universality.

\section{Drell-Yan production and Sudakov resummation}

Drell-Yan (DY) production, apart from its phenomenological significance, 
is theoretically interesting, 
because one can kinematically realize the situation of a process with 
two hard scales. In the following, we consider the partonic DY cross 
section $\hat{\sigma}_{DY}$ 
(after collinear subtractions) in the region $Q\gg Q (1-z)\gg 
\Lambda$, where $z=Q^2/\hat{s}$, $Q^2$ being the mass of the DY pair and 
$\hat{s}$ the partonic cms energy. The second scale $Q (1-z)$ can be 
identified with the energy available to parton emission into the final 
state. Since $Q (1-z)\ll Q$, these partons are referred to as soft, 
although they are not soft in terms of the QCD scale $\Lambda$.  
Taking moments in $z$ (roughly, this replaces $1/(1-z)$ by $N$), one 
obtains two powers of $\ln N$ for each power of $\alpha_s\equiv
\alpha_s(Q)$, so that the actual expansion parameter of the hard 
cross section is $\alpha_s\ln^2 N$. Thus, in higher orders, one has two 
sources of large corrections, Sudakov logarithms, related to the 
scale $Q (1-z)$ and renormalon factorials, related to the scale $\Lambda$. 
One may ask how this complication affects the arguments that lead 
to the identification of power corrections through renormalons.

This question has been addressed in Refs.~\cite{CON94,KOR95,BBDY}. 
Starting from 
\begin{equation}
\ln \hat{\sigma}_{DY}(N) = \frac{2C_F}{\pi}\!\!
\int\limits_0^1\!\!dz\frac{z^N-1}{1-z}\!\!\!\!\!
\int\limits_{Q^2(1-z)}^{Q^2(1-z)^2}\!\!\!\!
  \frac{dk^2_\perp}{k^2_\perp}\,\alpha_s(k_\perp)\,,
\label{LLA}
\end{equation}
which resums all leading logs $\ln N\,(\alpha_s\ln N)^k$ in the DIS 
scheme, one finds 
\cite{CON94,KOR95} that the integral has a renormalon ambiguity of 
order $N\Lambda/Q$ from the region of large $z$. However, the 
corresponding $n!$ occurs in far subleading logarithms, beyond the 
accuracy to which (\ref{LLA}) was derived. Keeping all subleading 
logarithms essentially implies that the hard cross section is evaluated 
exactly. In Ref.~\cite{BBDY} this has been done in the approximation 
of a single chain, interpreted as an approximation to the logarithm 
of the cross section. Using the equivalence of this approximation 
with taking an explicit IR cut-off, we choose, for illustration, a 
lower cut-off $\mu$ on the emitted gluon energy. Omitting all terms that 
can not give rise to a $\mu/Q$-correction (which allows us to ignore 
virtual corrections and collinear subtractions), one 
obtains instead of the right hand side of (\ref{LLA})
\begin{equation}
\label{DYphasespace}
\frac{2C_F\alpha_s}{\pi}\!\!\!\!\!\!
\int\limits_0^{1-2\mu/Q} \!\!\!\!\!\!\! dz\, z^{N-1}
\!\!\!\!\!\!\!\!\!\!\!\int\limits_{\mu^2}^{Q^2(1-z)^2/4}
\!\!\!\!\!\!
\frac{d k^2_\perp}{k^2_\perp}\frac{1}
{\sqrt{(1-z)^2-4k^2_\perp/Q^2}}\,
\end{equation}
which reduces to the structure of (\ref{LLA}) in the double-logarithmic, 
soft-collinear limit, if $k_\perp$ is set to zero in the square root. 
In this limit, consistent with the previous result, the integral 
contains a $\mu/Q$-term in the expansion in small $\mu$. However, 
the $k_\perp$-term is crucial and can not be neglected precisely in the 
region $z\to 1$, where the $\mu/Q$-term originates from. Keeping 
$k_\perp$ in the square root, one finds that $1/Q$-corrections are 
absent and that the leading power correction is of order $(N\mu/Q)^2$. 
The cancelation of the $1/Q$-term emphasizes that leading power 
corrections stem from soft gluons, but small angle (collinear) and 
large angle ($k_\perp\sim k_0\sim Q(1-z)$) emission are both important. 
\cite{BBDY} This might appear surprising, because large angle soft 
emission is usually considered suppressed, leading to angular ordering 
in parton cascades. But only logarithmic enhancements of matrix 
elements cancel at large angles and no conclusion follows for 
power corrections. 

Returning to resummation of Sudakov logarithms, we conclude that 
there is no direct connection between resummation of $\ln N$ terms 
and power corrections, which are `buried' among an infinite number 
of subleading logs. Roughly speaking, this is so, because 
$Q\gg Q (1-z)\gg \Lambda$ and Sudakov resummation is concerned only 
with the first inequality, while power corrections are associated with 
the smallest scale $\Lambda$. 
Nevertheless, one would like to formulate the 
resummation procedure in such a way that it does not introduce 
stronger power corrections than required. Given the hierarchy 
of scales, it is useful to think about the problem from the 
effective field theory point of view, which deals with scales 
sequentially. Thus, one would first `integrate out' momenta 
larger than $Q (1-z)$ and deal with Sudakov resummation in this 
step. At that lower scale, we can consider power corrections, which 
then appear as the ratio $\Lambda/(Q (1-z))$ ($N\Lambda/Q$ in 
moment space). 

The effective fields for fast quarks interacting with soft gluons 
are expressed as eikonal or Wilson lines. When $N\gg 1$, the 
hard Drell-Yan cross section (omitting for simplicity collinear 
subtractions) factorizes \cite{STE87,KM} as 
$\hat{\sigma}_{DY}(N,Q) = H_{DY}(Q,\mu)\,S_{DY}(Q/N,\mu)$, 
where the scales $Q$ and $Q/N$ are separated. The `soft part' 
$S$ satisfies a renormalization group equation in $\mu$ that 
can be used to sum logarithms in $N$, because $S$ depends only 
on the single dimensionless ratio $Q/(N\mu)$. The solution to 
the RGE is expressed as
\begin{eqnarray}
\label{eff}
\hat{\sigma}_{DY} &=& H_{DY}(Q)\cdot S_{DY}(1,\alpha_s(Q/N))\cdot  
\nonumber\\
&& \hspace*{-1.5cm}
\exp\Bigg(\,\int\limits_{Q^2/N^2}^{Q^2}\!\!\frac{d k_t^2}{k_t^2} 
\bigg[\Gamma_{eik}(\alpha_s(k_t))\ln\frac{k_t^2 N^2}{Q^2}
\nonumber\\
&&\hspace*{-1.5cm}\,+\Gamma_{DY}(\alpha_s(k_t))\bigg]\Bigg)\,.
\end{eqnarray}
The three factors on the r.h.s. correspond to coefficient function, 
matrix element and anomalous dimension factor in the effective 
field theory language. The analysis of Ref.~\cite{BBDY} shows that 
in the $\overline{\mbox{MS}}$ scheme the universal eikonal anomalous 
dimension and DY specific anomalous dimension $\Gamma_{DY}$ are 
analytic functions for small $\alpha_s$, so that the exponential 
factor does not contain any renormalons (power corrections) at all. 
These enter only through the boundary conditions at the lower and 
higher scale and turn out to be $(N\Lambda/Q)^2$ as stated before. 
Notice also that when $N>Q/\Lambda$, the exponent becomes 
ill-defined. For such large moments, the language of power 
corrections looses its meaning ($N \Lambda/Q\sim 1$), the energy 
of soft gluons becomes of order $\Lambda$ and any short-distance 
expansion fails.

Whether the absence of $1/Q$-corrections for Drell-Yan production 
persists beyond the approximation of a single chain 
(one-gluon emission), is an unsolved problem. In Ref.~\cite{AZ}, 
the cancelation at leading order has been reproduced as a consequence 
of the KLN and Low theorems and it has been argued to hold at the 
level of two-gluon emission as well. On the other hand, in the 
language of Wilson lines, emphasized in Refs.~\cite{KOR95,KOR}, 
a $1/Q$-correction could be naturally accommodated by a certain 
operator constructed from Wilson lines, although it vanishes at 
leading order. If, as suspected in Ref.~\cite{KOR}, Glauber gluons 
constitute a new potential source of $1/Q$-corrections at higher 
orders, the validity of the eikonal approximation and 
Wilson line treatment to power-like 
accuracy would have to be re-examined. In principle, the 
possibility exists \cite{BBDY} that the cancelation of 
$1/Q$ terms occurs between (\ref{eff}) and terms dropped in 
(\ref{eff}), although there is no indication of it at leading 
order. Since, as explained above, the 
problem of resummation is disconnected from the problem of power 
corrections, one should be able to establish equivalence of 
the renormalon analysis for the DY cross section with 
the analysis of power corrections in terms of multi-parton correlation 
functions.\cite{QIU91}  The present discussion indicates that 
the optimal language for problems with two scales might yet have to be  
found.

\section{The Sudakon: $x$-space vs. $N$-space resummation} 

As pointed out in Ref.~\cite{tt3}, factorial divergence, not related 
to renormalons, can appear 
when one converts resummed distributions in moment space 
($N$-space) back to `real' space ($x$-space; note that $x$ replaces 
$z$ in the previous section). Typically, a physical quantity is given 
as an integral
\begin{equation}
\label{incl}
\sigma(\tau)=\int\limits_0^1 d x\,W(\tau/x)\,\hat{\sigma}(x)\,,
\end{equation}
where $\hat{\sigma}$ could be the partonic Drell-Yan cross section 
-- in which case $W$ is the parton luminosity and $\tau=Q^2/s$ -- 
or the thrust distribution, or the lepton energy distribution 
in semileptonic $B\to X_u l\nu$ decay, for example. Let $x$ be a 
generic variable, such that $x\to 1$ corresponds to the soft gluon 
region. If the weight function $W$ constrains $x$ to the region of 
large $x$, but such that $1-x$ is still large compared to $\Lambda/Q$, 
where $Q$ is the generic hard scale, the perturbative expansion 
of $\sigma(\tau)$ contains the usual renormalon factorials and the 
corresponding poles in the Borel transform.

Consider now the double logarithmic approximation with fixed coupling 
$\alpha_s$, in which all $(\alpha_s\ln^2 N)^k$ have been resummed in 
moment space. It is perfectly consistent to perform the inverse Mellin 
transformation to $x$-space in the same approximation. Then 
$\hat{\sigma}(x)\approx \exp(c\alpha_s\ln^2(1-x))$ with some constant 
$c$. Expanding $\sigma(\tau)$ in $\alpha_s$, one finds the 
divergent series
\begin{eqnarray}
\label{integral}
&&\hspace*{-0.3cm}
\int\limits_0^1 d x\,W(\tau/x)\,\exp(c\alpha_s\ln^2(1-x)) 
\nonumber\\
&&\hspace*{-0.3cm}\sim \sum_n F_n(\tau)\,(4 c)^n n! \,\alpha_s^{n+1}\,
\end{eqnarray}
independent of $W$, as long as it does not depend exponentially on 
$x$. 
In practical cases $c$ is such that this series diverges much faster 
than expected from renormalons. Continuing the tradition of 
misnomers, I call the corresponding singularity in the Borel plane 
a Sudakon pole. It is unphysical and appears, because in the 
process of resummation, one has dropped terms, which after integration 
over $x$ give equally large contributions and cancel the 
singularity. 

On physical grounds, one expects Sudakov suppression, so $c$ is 
negative. Then the integral in (\ref{integral}) can just be done 
and there is no reason to re-expand it in $\alpha_s$. Even in this 
case, however, resummation has modified the analytic structure 
of the Borel transform, although the spurious pole is Borel summable. 
While $c$ is indeed negative for the thrust distribution or lepton energy 
distribution in $B$ decay, mentioned above, it is in fact 
positive for Drell-Yan production (and other hadronic collisions) 
in the conventional subtraction schemes, because the subtracted 
product of partonic structure functions shows stronger Sudakov 
suppression than the partonic Drell-Yan cross section. In this case, 
the integral in (\ref{integral}) does not exist and must be 
defined by truncating the divergent expansion at its minimal term. 
If, as usual, one interprets the size of the minimal term 
as an uncertainty in defining 
the integral, one finds that this uncertainty is of order 
$(\Lambda/Q)^{\beta_0/4 c}$. As $\beta_0/(4 c)$ can be much smaller 
than one \cite{tt3}, this uncertainty is large. Alternatively, the 
integral can be defined by excluding the region of large $x$. Truncation 
of the series is equivalent to a cut-off in $x$ that corresponds to 
excluding gluons with energy larger than $\Lambda\cdot (Q/\Lambda)^{1-
\beta_0/(2 c)}$ from a perturbative treatment. This cut grows with 
$Q$ and is much larger than the expected limit of order 
$\Lambda$ for perturbative gluons. In fact, the energy cut corresponds 
to the position 
of the Sudakov peak of the $x$-distribution, if $c$ were negative.

Although the sketched procedure is consistent from the point of view 
of summing logarithms in $1-x$, it is desirable to formulate resummation 
for integrated quantities as in (\ref{incl}) such that factorial 
behaviour in their expansions is consistent with the expected 
renormalon structure. In Ref.~\cite{tt3} it is proposed to perform 
the inverse Mellin transform of the 
double-log resummed $\hat{\sigma}(N)$ back to 
$x$-space exactly, keeping all subleading logarithms. 
In this procedure, resummation 
by itself does not introduce any factorial behaviour, as suggested also 
by the discussion of Sect.~3. Exact evaluation of the inverse 
Mellin transformation keeps exactly those subleading logarithms, which 
are necessary to cancel the unphysical Sudakon pole. This, together 
with the discussion of $1/Q$-corrections before, emphasizes that often 
it is not sufficient to perform resummations that are consistent to 
a certain logarithmic accuracy. Different treatments of subleading 
logarithms can result in numerically important different constant terms. 
Renormalon considerations can help to decide whether such terms 
are spurious or not.

\section*{Acknowledgments}

I am grateful to Volodya Braun for many discussions related to the 
subject of this article and Arthur Hebecker for comments 
on the text.

\end{document}